\documentstyle[prl,aps]{revtex}
\begin{document}
\draft
\twocolumn[\hsize\textwidth\columnwidth\hsize\csname @twocolumnfalse\endcsname
\title{Squeezed Phonon States: \\
Modulating Quantum Fluctuations of Atomic Displacements}
\author{Xuedong Hu and Franco Nori}
\address
{Department of Physics, The University of Michigan, Ann Arbor, Michigan
48109-1120}
\date{\today}
\maketitle
\begin{abstract}
We study squeezed quantum states of phonons, which allow the
possibility of modulating the quantum fluctuations of atomic
displacements below the zero-point quantum noise level of coherent
phonon states.  We calculate the corresponding expectation values and
fluctuations of both the atomic displacement and the lattice amplitude
operators, and also investigate the possibility of generating squeezed
phonon states using a three-phonon parametric amplification process
based on phonon-phonon interactions.  Furthermore, we also propose a
detection scheme based on reflectivity measurements.
\end{abstract}

\pacs{PACS numbers: 42.50.Lc, 42.50.Dv, 05.40.+j, 63.90.+t}
%
%
\vskip2pc]
\narrowtext

{\it Introduction.}---Photon squeezed states have attracted much
attention during the past decade \cite{special}.  These states are
important because they can achieve lower quantum noise than the
zero-point fluctuations of the vacuum or coherent states. Thus they
provide a way of manipulating quantum fluctuations and have a promising
future in different applications ranging from optical communications to
gravitational wave detection \cite{special}.  Indeed, squeezed states
are currently being explored in a variety of non-quantum-optics
systems, including classical squeezed states \cite{Rugar}.  Here we
want to study the properties of {\it phonon} squeezed states and
explore the possibility of generating these states through
phonon-phonon interactions.  After briefly presenting the quantum
mechanical description of various kinds of phonon states, we study a
simple model for generating phonon squeezed states, in which analytical
results can be obtained \cite{xhu}.  We also propose a scheme for
detecting this squeezing effect.

In most macroscopic situations, a classical description is adequate.
However, the quantum fluctuations of a phonon system can be dominant at
low enough temperatures.  Indeed, a recent study shows that quantum
fluctuations in the atomic positions can influence observable
quantities (e.g., the Raman line-shape) \cite{fnori} even when
temperatures are not very low.

An experimentally observable quantity for a phonon system is the real
part of the Fourier transform of the atomic displacement:  $Re\left(
u_{\alpha}({\bf q}) \right) = \sum_{\lambda} \sqrt{ {\hbar}/{8m
\omega_{{\bf q}\lambda}} }$ $\{ U^{\lambda}_{{\bf q} \alpha} ( b_{{\bf
q}\lambda} + b_{{-\bf q}\lambda}^{\dagger} ) + U^{\lambda \; *}_{{\bf
q} \alpha} ( b_{{-\bf q}\lambda} + b_{{\bf q}\lambda}^{\dagger} ) \}$
\cite{phonon}.  For simplicity, hereafter we will drop the branch
subscript $\lambda$, assume that $U_{{\bf q} \alpha}$ is real, and
define a ${\bf q}$-mode dimensionless lattice amplitude operator:
$u(\pm {\bf q}) = b_{\bf q} + b^{\dagger}_{-\bf q} + b_{-\bf q} +
b^{\dagger}_{\bf q}$.  This operator contains essential information on
the lattice dynamics, including quantum fluctuations.  It is the phonon
analog of the electric field in the photon case.

{\it Phonon Vacuum and Number States.}---When no phonon is excited, the
crystal is in the phonon vacuum state $|0\rangle$.  The eigenstates of
the harmonic phonon Hamiltonian are number states which satisfy $b_{\bf
q}|n_{\bf q}\rangle = \sqrt{n_{\bf q}} |n_{\bf q}-1\rangle$.  The
phonon number and the phase of atomic vibrations are conjugate
variables.  Thus, due to the uncertainty principle, the phase is
arbitrary when the phonon number is certain, as it is the case with any
number state $|n_{\bf q}\rangle$.  Therefore, the expectation values of the
atomic displacement $\langle n_{\bf q}|u_{i\alpha}|n_{\bf q} \rangle$
and ${\bf q}$-mode lattice amplitude $\langle n_{\bf q}|u(\pm {\bf
q})|n_{\bf q} \rangle$ vanish due to the randomness in the phase of the
atomic displacements.

{\it Phonon Coherent States.}---A single-mode (${\bf q}$) phonon
coherent state is an eigenstate of a phonon annihilation operator:
$b_{\bf q}|\beta_{\bf q}\rangle = \beta_{\bf q} |\beta_{\bf q} \rangle$
\cite{coherent}.  It can also be generated by applying a phonon
displacement operator $D_{\bf q}(\beta_{\bf q})$ to the phonon vacuum
state $|\beta_{\bf q} \rangle = D_{\bf q}(\beta_{\bf q})$ $|0\rangle =
\exp(\beta_{\bf q} b_{\bf q}^{\dagger} - \beta_{\bf q}^* b_{\bf
q})|0\rangle = \exp(-|\beta_{\bf q}|^2/2) \sum_{n_{\bf q}=0}^{\infty}
\beta_{\bf q}^{n_{\bf q}}$ $|n_{\bf q}\rangle /\sqrt{n_{\bf q}!}\,$.
Thus it can be seen that a phonon coherent state is a phase coherent
superposition of number states.  Moreover, coherent states are a set of
minimum-uncertainty states which are as noiseless as the vacuum state
\cite{fluc}.  Coherent states are also the quantum states that best
describe the classical harmonic oscillators \cite{Gardiner}.

{\it Phonon Squeezed States.}---In order to reduce quantum noise to a
level below the zero-point fluctuation level, we need to consider
phonon squeezed states.  Quadrature squeezed states are generalized
coherent states \cite{Loudon}.  Here ``quadrature'' refers to the
dimensionless coordinate and momentum.  Compared to coherent states,
squeezed ones can achieve smaller variances for one of the quadratures
during certain time intervals and are therefore helpful for decreasing
quantum noise.

A single-mode quadrature phonon squeezed state is generated from a
vacuum state as $|\alpha_{\bf q} , \xi \rangle = D_{\bf q}(\alpha_{\bf
q} ) S_{\bf q}(\xi )$ $|0\rangle$; a two-mode quadrature phonon
squeezed state is generated as $|\alpha _{{\bf q}_1}, \alpha _{{\bf
q}_2}, \xi \rangle = D_{{\bf q}_1} (\alpha _{{\bf q}_1})D_{{\bf q}_2}
(\alpha _{{\bf q}_2}) S_{{\bf q}_1, {\bf q}_2} (\xi) |0\rangle$.
Here $D_{\bf q}(\alpha_{\bf q})$ is the coherent state displacement
operator with $\alpha_{\bf q} = |\alpha_{\bf q}| e^{i\phi}$, $S_{\bf q}
(\xi ) = \exp (\xi^*b_{\bf q}^2/2 - \xi b_{\bf q}^{\dagger \, 2}/2 )$
and $S_{{\bf q}_1, {\bf q}_2} (\xi ) = \exp(\xi^* b_{{\bf q}_1} b_{{\bf
q}_2} - \xi b_{{\bf q}_1}^{\dagger} b_{{\bf q}_2}^{\dagger})$ are the
single- and two-mode squeezing operator \cite{squeeze}, and $\xi=r
e^{i\theta}$ is the complex squeezing factor with $r \geq 0$ and $0
\leq \theta < 2\pi$.  The two-mode phonon quadrature operators have the
form $X({\bf q},{-\bf q}) = (b_{\bf q} + b_{\bf q}^{\dagger} + b_{-\bf
q} + b_{-\bf q}^{\dagger})/2^{3/2} = 2^{-3/2}u(\pm {\bf q})$ and
$P({\bf q},{-\bf q}) = (b_{\bf q} - b_{\bf q}^{\dagger} + b_{-\bf q} -
b_{-\bf q}^{\dagger})/(2^{3/2}i)$.

We have considered two cases where squeezed states were involved in
modes $\pm {\bf q}$.  In the first case, the system is in a two-mode
($\pm {\bf q}$) squeezed state $|\alpha_{\bf q}, \alpha_{-\bf q}, \xi
\rangle$, ($\xi = re^{i\theta}$), with fluctuations $\langle(\Delta
u(\pm {\bf q}))^2\rangle_{\rm sq} = 2(e^{-2r} \cos^2 \frac{\theta}{2} +
e^{2r} \sin^2 \frac{\theta}{2})$.  In the second case, the system is in
a single-mode squeezed state $|\alpha_{\bf q}, \xi \rangle$,
($\alpha_{\bf q} = |\alpha_{\bf q}|e^{i\phi}$), in the first mode and
an arbitrary coherent state $|\beta_{-\bf q} \rangle$ in the second
mode.  The fluctuation is now $1 + e^{-2r}\cos^2 (\phi +
\frac{\theta}{2}) + e^{2r} \sin^2 (\phi + \frac{\theta}{2})$.  In both
of these cases, $\langle(\Delta u(\pm {\bf q}))^2\rangle_{\rm sq}$ can
be smaller than in coherent states.

{\it Phonon Parametric Process.}---Now we propose a scheme to generate
phonon squeezed states \cite{general,numcons}.  This scheme is based on
a ``phonon'' parametric amplification process (e.g., the decaying
process: LO phonon $\rightarrow$ two LA phonons, where LO refers to
Longitudinal Optical and LA to Longitudinal Acoustic), which in turn is
based on three-phonon interactions.  Typically, three-phonon
interactions are the dominant anharmonic processes in a phonon system
and the lowest order perturbation to the harmonic Hamiltonian.  We will
neglect all the higher order interactions because they are generally
much weaker than the third-order ones.  For all parametric processes,
the pump wave (of phonons in this case) must be very strong because the
generic physical processes inside parametric amplifiers are generally
nonlinear and weak.  This pumping process can be realized by using two
pulse lasers to illuminate a crystal.  With appropriate laser
frequencies and directions, coherent LO phonons of the pump mode at the
Brillouin-Zone-center can be generated through, for example, stimulated
Raman scattering (provided that the pump mode is Raman active), as
discussed, e.g., in Refs.~\cite{Yariv,review}.

The Hamiltonian for the whole process initiated by the Raman scattering
is (see Fig.~\ref{fig1})
\begin{eqnarray}
H_{\rm param} & = & H_0 + H_{\rm Raman} + H_{\rm anh}  \\
H_0 & = & \hbar\omega_{{\bf k}_1}a_{{\bf k}_1}^{\dagger}a_{{\bf k}_1}
	+ \hbar\omega_{{\bf k}_2}a_{{\bf k}_2}^{\dagger}a_{{\bf k}_2} +
	\sum_{\bf q}\hbar\omega_{\bf q}b_{\bf q}^{\dagger}b_{\bf q}
	\nonumber \\
H_{\rm Raman} & = & \eta a_{{\bf k}_1} a_{{\bf k}_2}^{\dagger}
	b_{{\bf q}_p}^{\dagger} + \eta^* a_{{\bf k}_1}^{\dagger}
	a_{{\bf k}_2} b_{{\bf q}_p}  \nonumber \\
H_{\rm anh} & = & \lambda_{{\bf q}_s{\bf q}_i} b_{{\bf q}_p}
	b_{{\bf q}_s}^{\dagger} b_{{\bf q}_i}^{\dagger} + \lambda_{{\bf
	q}_s{\bf q}_i}^* b_{{\bf q}_p}^{\dagger} b_{{\bf q}_s} b_{{\bf
	q}_i} \nonumber \\
	& & + \sum_{{\bf q}'{\bf q}''}(\lambda_{{\bf q}'{\bf q}''}
	b_{{\bf q}_p} b_{{\bf q}'}^{\dagger} b_{{\bf q}''}^{\dagger} +
	\lambda_{{\bf q}'{\bf q}''}^* b_{{\bf q}_p}^{\dagger} b_{{\bf
	q}'} b_{{\bf q}''}) \,. \nonumber
\end{eqnarray}
Here $a$ ($b$) refer to photon (phonon) operators.  The higher-
(lower-) energy incident photon mode is labeled by ${\bf k}_1$ (${\bf
k}_2$).  Notice that the lower energy photon mode is generally called
Stokes mode in the context of Raman scattering.  The sums over ${\bf
q}'$ and ${\bf q}''$ in $H_{\rm anh}$ represent decay channels other
than the special one with acoustic signal and idler modes.

We now consider two mean field averages in order to simplify an
otherwise analytically intractable problem.  The first mean field is
over the photons.  The photons in the incident modes ${\bf k}_1$ and
${\bf k}_2$ (often denoted by ``laser'' and ``Stokes'' light) originate
from two lasers.  As long as these two incident laser modes are not
strongly perturbed by the Raman scattering process, we can treat both
of these incoming photon states as coherent states $|\alpha_{{\bf k}_1}
\, e^{-i\omega_{{\bf k}_1}t} \rangle$ and $|\alpha_{{\bf k}_2} \,
e^{-i\omega_{{\bf k}_2}t} \rangle$, and perform a mean field average
over them.  The second mean field average is over the LO pump mode
phonons.  Since phonons produced by coherent or stimulated Raman
scattering are initially in coherent states, we denote this pump mode
phonon coherent state as $|\beta_0 (t) \rangle$, with $\langle \beta_0
(t) | b_{{\bf q}_p} |\beta_0 (t) \rangle = \beta_0 (t)$.  Since these
LO phonons are in coherent states, the results from the average over
the pump mode phonons are c-numbers with a well-behaved
time-dependence.  Now we drop all the c-number terms because they will
not affect our results.  In addition, we will also drop all the phonon
modes involved in the decay channels other than the special one
consisting of the signal modes, considering them only weakly coupled to
the pump mode; i.e., we assume $\lambda_{{\bf q}'{\bf q}''} \ll
\lambda_{{\bf q}_s{\bf q}_i}$.  The Hamiltonian now becomes
\begin{eqnarray}
H_{\rm param}^{\prime} & = & \hbar \omega_{{\bf q}_s}
b^{\dagger}_{{\bf q}_s} b_{{\bf q}_s} + \hbar \omega _{-{\bf q}_s}
b^{\dagger}_{-{\bf q}_s} b_{-{\bf q}_s}  \nonumber \\
& & \hspace{-0.4in} + \lambda_{{\bf q}_s,{-\bf q}_s} \, \beta_0(t) \,
b^{\dagger}_{{\bf q}_s} b^{\dagger}_{-{\bf q}_s} + \lambda_{{\bf
q}_s,{-\bf q}_s}^* \, \beta_0^*(t) \, b_{{\bf q}_s} b_{-{\bf q}_s},
\label{eqn:3-para}
\end{eqnarray}
where $\beta_0 (t)$ is the coherent amplitude of the pump mode
phonons.  We use $H_0 + H_{\rm Raman}$ to determine $\beta_0 (t)$, and
then substitute it back into $H_{\rm param}$ to obtain $H_{\rm
param}^{\prime}$.  Here we have implicitly assumed that the Raman
scattering process is stronger than the anharmonic scattering.
According to our previous discussion \cite{squeeze}, the two-mode LA
phonon system will evolve into a two-mode squeezed state $|\alpha_{{\bf
q}_s}, \alpha_{{-\bf q}_s}, \xi(t) \rangle$ from an initial coherent or
vacuum state, with a squeezing factor of
\begin{equation}
\xi(t) = \frac{i}{\hbar}\int_{0}^t \lambda_{{\bf q}_s,{-\bf
q}_s} \, \beta_0 (\tau) \, e^{2i\omega_{{\bf q}_s} \tau} d\tau = \,.
\end{equation}
which is only valid in the very short time limit (i.e., small $t$).

In summary, we have just considered generating two-mode LA phonon
squeezed states $|\alpha_{{\bf q}_s}, \alpha_{{-\bf q}_s}, \xi(t)
\rangle$ by using the three-phonon anharmonic interaction
\cite{2Raman}.  The higher-energy LO phonon mode, which is called the
``pump'' mode, is driven into a coherent state through stimulated Raman
scattering.  This mode in turn is used as a pump in the parametric
amplification process involving itself and the two lower-energy LA
phonon modes ($\pm {\bf q}_s$), the signal and the idler.  Both of
these modes can here be called ``signal'' because the ``idler'' mode is
not really ``idle''; indeed, it is actively involved in the squeezing
process.  In conclusion, we have shown that the LA phonons in the two
signal modes ($\pm {\bf q}_s$) are in a two-mode squeezed state if
$(i)$ the LO pump mode is in a coherent state and $(ii)$ we can neglect
the other decay channels.

{\it Detection Schemes.}---It is possible to directly detect a
single-mode phonon squeezed state with phonon counters \cite{phononopt}
such as superconducting tunnel junction bolometers and vibronic
detectors.  The signature of a single-mode squeezed state is a
sub-Poissonian phonon number distribution in that mode.  However, these
phonon counters are either wide-band, or have low efficiency.
Therefore, direct detection might not be the best method to detect
squeezing effect.

Phase-sensitive schemes such as homodyne and heterodyne detectors are
most often used to detect photon squeezed states because of their
ability to lock phase with the electric field of the measured state
\cite{Loudon}.  There appears to be no available phase-sensitive
detection method for phonons.  A promising candidate might be measuring
the intensity of a reflected probe light \cite{review}.  This method
has already been used to detect phonon amplitudes, since the
reflectivity is linearly related to the atomic displacements in a
crystal.  The value of the lattice amplitude operator can be extracted
by making a Fourier analysis on the sample reflectivity.  If squeezing
should happen, its effect will be contained in the Fourier components
of the intensity of the reflected light.  In this manner the
information on the squeezing effect in the phonons is also carried by
the reflected light in the form of squeezing of the photon intensity.
We can then use a standard optical detection method to determine
whether the related light is squeezed or not.  One shortcoming of this
method is that it is not direct.  In the measurement there can be noise
added into the signal, such as the intensity fluctuation of the
original probe light, the efficiency for the reflected light to pick up
the signals in the phonons, etc.  Needless to say, further research
needs to be done on how to realize this phase-sensitive detection
scheme, and we hope that our initial proposals stimulate further
theoretical and experimental work on this problem.

{\it Discussion.}---Phonon squeezing depends on the absolute value $r$
and also on the phase $\theta$ of the squeezing factor $\xi(t) =
re^{i\theta}$.  More explicitly, $\langle(\Delta u(\pm {\bf
q}))^2\rangle_{\rm sq} = 2 \left( e^{-2r} \cos^2{\frac{\theta}{2}} +
e^{2r} \sin^2{\frac{\theta}{2}} \right)$.  Only when $\theta$ is close
to $0$ is noise suppressed in the lattice amplitude operator.  This
means that in order to suppress the noise, the squeezing factor
$\xi(t)$ has to have a dominant positive real part so that $\cos\theta
> \tanh r$.  The squeezing factor obtained from the three-phonon
process is $\xi(t) = \frac{i}{\hbar}\int_0^t \lambda \, \alpha(\tau) \,
e^{i(\omega_s + \omega_i)\tau} d\tau$, where the real number $\lambda$
is the strength of the interaction and $\alpha$ is the amplitude of the
phonon coherent state in the pump mode.  From this expression for
$\xi(t)$ we can see that the squeezing effect only appears during
certain time intervals.  If $\alpha(t)$ does not depend on time or has
a periodic dependence on time, squeezing will be periodic in time,
which makes phase-sensitive detection easier to achieve.

To make the above schemes work, some noise problems have to be
overcome.  First, any attempt to generate or detect squeezed states
should be at low temperatures to avoid thermal noise in the crystal.
For instance, the excitation energy of a $10$THz optical phonon
corresponds to a temperature of about $100$K.  Therefore, the
experiment might have to be carried out at a temperature well below
$100$K, such as $10$K or lower.  Second, the fluctuations in the laser
intensity and in the interaction between the laser and the crystal has
to be very small, so that they will not suppress the noise reduction
process in the squeezing effect.  Indeed, one of the possible ways to
reduce the noise coming from the laser beam is to use a beam of
squeezed photons.  Finally, the incoherence in the procedure itself has
to be minimized.  For example, the finite lifetime of pump mode phonons
does not favor the generation of squeezed states because it gives rise
to an additional noise in the intensity of the mode.  Therefore, we
need long lifetime LO phonons, which can be realized in, for instance,
materials with weak anharmonic interactions and low concentration of
isotopic defects (e.g., diamond).  Furthermore, here we have studied a
discrete-mode phonon parametric process.  A continuous-mode model will
be discussed elsewhere.

{\it Conclusions.}---We have investigated the dynamics and quantum
fluctuation properties of phonon squeezed states.  In particular, we
calculate the experimentally observable time evolution and fluctuation
of the lattice amplitude operator $u(\pm {\bf q})$, and show that
$\langle u(\pm {\bf q}) \rangle_{\rm sq}$ is a sinusoidal function of
time, while $\langle (\Delta u(\pm {\bf q}))^2 \rangle_{\rm sq}$ is
periodically smaller than the vacuum and coherent state value $2$.  In
other words, phonon squeezed states are periodically quieter than the
vacuum state.  We have discussed one particular approach to generate
phonon squeezed states.  This approach is based on a three-phonon
process where the higher energy optical phonon mode is coherently
pumped.  We show that the two lower-energy acoustic phonon modes can be
in a two-mode phonon quadrature squeezed state given appropriate
initial conditions.  We achieve this by dealing separately with $(i)$
the optical excitation of the pump mode optical phonons and $(ii)$ the
anharmonic scattering of the pump mode phonons into the lower-energy
acoustic phonons.  We have also briefly analyzed a potential detection
method of phonon squeezed states.  Experiments in quantum optics
indicate that phase-sensitive methods---such as homodyne
detection---are the best in detecting photon squeezed states.
Therefore, we have proposed a detection scheme based on a reflected
probe light and an ordinary phase-sensitive optical detector.

Like in the photon case \cite{special}, the experimental realization of
phonon squeezed states might require years of work after its initial
proposal.  We hope that our effort will lead to more theoretical and
experimental explorations in the area of phonon quantum noise
modulation.

It is a great pleasure for us to acknowledge useful conversations with
Saad Hebboul, Hailin Wang, Roberto Merlin, Duncan Steel and especially
Shin-Ichiro Tamura.


\begin{figure}
\caption{A schematic diagram of a three-phonon parametric process.
Here (a) refers to a stimulated Raman scattering and (b) to a
three-phonon anharmonic scattering process.  The subscript ${\bf k}_1$
(${\bf k}_2$) refers to the higher- (lower-) energy incident coherent
photons.  The arrows in the diagram illustrate the directions of the
photon and phonon momentum vectors.  A typical process is as follows: a
photon in mode ${\bf k}_1$ interacts with the phonon system and emits
one LO phonon in the pump mode of frequency $\omega_{p}$, while the
photon itself is scattered into mode ${\bf k}_2$; the generated pump
mode LO phonon proceeds in the crystal, interacts with the lower-energy
phonon modes through the three-phonon interaction, and eventually
splits into two LA phonons in modes $s$ and $s'$.  The latter can be
squeezed for appropriate initial states.  Notice that the pump mode LO
phonons have an almost-zero wave vector, so that the two lower-energy
LA phonon modes have nearly opposite wave vectors $\pm {\bf q}_s$.
Also notice that this figure is not to scale.  For the sake of clarity,
we have increased the angle between the lines, and drawn a longer line
for the pump mode phonon.}
\label{fig1}
\end{figure}


\begin{thebibliography}{99}
\bibitem{squeeze}  The squeezing operator $S_{{\bf q}_1, {\bf
q}_2}(\xi)$ can be produced by the following Hamiltonian: $H_{{\bf
q}_1, {\bf q}_2} = \hbar \omega_{{\bf q}_1} b^{\dagger}_{{\bf q}_1}
b_{{\bf q}_1} + \hbar \omega_{{\bf q}_2} b_{{\bf q}_2}^{\dagger}
b_{{\bf q}_2} + \zeta(t) b^{\dagger}_{{\bf q}_1} b^{\dagger}_{{\bf
q}_2} + \zeta^*(t)  b_{{\bf q}_1} b_{{\bf q}_2}$.  The time-evolution
operator has the form $U(t) = \exp (-i H_0 t/\hbar)$ $\exp [ \xi^*(t)
b_{{\bf q}_1}b_{{\bf q}_2} - \xi(t) b_{{\bf q}_1}^{\dagger} b_{{\bf
q}_2}^{\dagger} ]$, where $H_0 =  \hbar \omega_{{\bf q}_1}
b^{\dagger}_{{\bf q}_1} b_{{\bf q}_1} + \hbar \omega_{{\bf q}_2}
b^{\dagger}_{{\bf q}_2} b_{{\bf q}_2}$ and $\xi(t) =  \frac{i}{\hbar}
\int_{-\infty}^t \zeta(\tau) \, e^{i(\omega_{{\bf q}_1} + \omega_{{\bf
q}_2}) \tau} d\tau $.  Here $\xi(t)$ is the squeezing factor, and
$\zeta(t)$ is the strength of the interaction between the phonon system
and the external source; this interaction allows the generation and
absorption of two phonons at a time.
\bibitem{general} It is important to point out that the squeezing
effect is quite general.  In fact, any diagonalizable system can be
``squeezed'' in some manner by an appropriate choice of initial
states.  For example, consider a free oscillator $H = a^{\dagger}a$,
and define $d = \mu a + \nu a^{\dagger}$, where $\mu$ and $\nu$ are
complex scalars.  In order for $d$ to be a boson operator, $\mu$ and
$\nu$ must satisfy the relation $|\mu|^2 - |\nu|^2 = 1$.  Then the
Hamiltonian becomes $H = (|\mu|^2 + |\nu|^2)d^{\dagger}d - (\mu \nu
d^{\dagger \; 2} + \mu^* \nu^* d^2) + |\nu|^2$.  The generalized
``displacement'' $d + d^{\dagger}$ is a mixture of the original
displacement $(a+a^{\dagger})/\sqrt{2}$ and momentum
$-i(a-a^{\dagger})/\sqrt{2}$.  If we are interested in a quantity such
as $d + d^{\dagger}$, and it is observable, the squeezing effect can be
obtained when the initial state is a coherent state of mode ``$d$'',
i.e., $d |\delta \rangle = \delta | \delta \rangle$.  On the other
hand, such a hybrid quantity $d + d^{\dagger}$ might not be physically
interesting and/or observable.  In other words, the squeezing effect is
only relevant when it is related to the observables.
\bibitem{2Raman} We have also studied a second-order Raman scattering
process.  Similar to the phonon parametric process, if the two incident
light beams are in coherent states, the acoustic phonons generated by
the second-order Raman scattering are in a two-mode squeezed state.  A
mean field average over the photons is also required in this case.
\end{thebibliography}
\end{document}